\documentclass[reprint, prd, superscriptaddress, longbibliography, tightenlines, nofootinbib, eqsecnum, floatfix, notitlepage,twocolumn]{revtex4-1}
\pdfoutput=1
\usepackage[utf8]{inputenc}
\usepackage{euscript}
\usepackage{epsfig}
\usepackage{graphics}
\usepackage{graphicx}
\usepackage{amsmath}
\usepackage{amssymb}
\usepackage{amsfonts}
\usepackage{bm}
\usepackage[usenames,dvipsnames,svgnames,table]{xcolor}
\usepackage{xspace}
\usepackage{times}
\usepackage{appendix}
\usepackage{lipsum}
\usepackage[nolist,nohyperlinks]{acronym}
\usepackage{float}
\usepackage{simplewick}
\usepackage{tabularx}
\usepackage{booktabs}
\usepackage{natbib, ifthen}
\usepackage{hyperref}
\usepackage{comment}
\usepackage{soul}
\usepackage[caption=false]{subfig}
\usepackage{orcidlink}

\newcommand{\vl}{\ensuremath{\boldsymbol{l}}\xspace}

\newcommand{\nlzero}{\ensuremath{N_L^{(0)}}\xspace}
\newcommand{\nlrdzero}{\ensuremath{\mathrm{RD{\text -}}N_L^{(0)}}\xspace}
\newcommand{\nlrdmapzero}{\ensuremath{\mathrm{RD{\text -}}N_L^{(0), \rm MAP}}\xspace}
\newcommand{\nlone}{\ensuremath{N_L^{(1)}}\xspace}
\newcommand{\nlzeroMAP}{\ensuremath{N_L^{(0), \rm MAP}}\xspace}
\newcommand{\nloneMAP}{\ensuremath{N_L^{(1), \rm MAP}}\xspace}

\newcommand{\Xdat}[0]{\ensuremath{ {X^{\rm dat}}}\xspace}% CMB data
\newcommand{\Xdatx}[0]{\ensuremath{ {X^{\rm dat}_{\times}}}\xspace}% CMB data
\newcommand{\cpp}[0]{\ensuremath{C^{\phi\phi}_L}\xspace}

\newcommand{\cppfid}[0]{\ensuremath{C^{\phi\phi, \rm fid}_L}\xspace}

\newcommand{\pMAP}[0]{\ensuremath{\phi^{\rm MAP}}\xspace}
\newcommand{\pMAPx}[0]{\ensuremath{\phi^{\rm MAP, \times}}\xspace}

\newcommand{\Cov}[0]{\ensuremath{\textrm{Cov}}\xspace} % Covariance matrix

\newcommand{\vL}[0]{\ensuremath{\boldsymbol{L}}\xspace}

\newcommand{\anorm}[0]{\ensuremath{\alpha}} % letter. Aim is to have to change this one only
\newcommand{\va}[0]{\ensuremath{\boldsymbol{\anorm}}\xspace} % vector field
\newcommand{\Da}[0]{\ensuremath{\mathcal D_{\va}}} % deflection operator at alpha
\newcommand{\Beam}[0]{\ensuremath{\mathcal B}\xspace}
\newcommand{\Cova}{\ensuremath{\textrm{Cov}_{\va}}} % data.
\newcommand{\Covai}{\ensuremath{\textrm{Cov}^{-1}_{\va}}} % data.

\newcommand{\tr}{\ensuremath{\text{Tr}}}

\newcommand{\gQD}[0]{\ensuremath{g^{\rm QD}}\xspace}
\newcommand{\gQDx}[0]{\ensuremath{g^{\rm QD, \times}}\xspace}
\newcommand{\gMF}[0]{\ensuremath{g^{\rm MF}}\xspace}

\graphicspath{ {./figures/} }

\begin{document}

\title{An iterative CMB lensing estimator minimizing instrumental noise bias}

\author{Louis Legrand\,\orcidlink{0000-0003-0610-5252}}\email{ll783@cam.ac.uk}
\affiliation{Department of Applied Mathematics and Theoretical Physics, University of Cambridge, Wilberforce Road, Cambridge CB3 0WA, United Kingdom}
\affiliation{Kavli Institute for Cosmology, Cambridge, Madingley Road, Cambridge CB3 OHA, United Kingdom}

\author{Blake Sherwin}
\affiliation{Department of Applied Mathematics and Theoretical Physics, University of Cambridge, Wilberforce Road, Cambridge CB3 0WA, United Kingdom}
\affiliation{Kavli Institute for Cosmology, Cambridge, Madingley Road, Cambridge CB3 OHA, United Kingdom}

\author{Anthony Challinor\,\orcidlink{0000-0003-3479-7823}}
\affiliation{Institute of Astronomy, Madingley Road, Cambridge CB3 OHA, United Kingdom}
\affiliation{Department of Applied Mathematics and Theoretical Physics, University of Cambridge, Wilberforce Road, Cambridge CB3 0WA, United Kingdom}
\affiliation{Kavli Institute for Cosmology, Cambridge, Madingley Road, Cambridge CB3 OHA, United Kingdom}

\author{Julien Carron}
\affiliation{Universit\'e de Gen\`eve, D\'epartement de Physique Th\'eorique et CAP, 24 Quai Ansermet, CH-1211 Gen\`eve 4, Switzerland}

\author{Gerrit~S. Farren}
\affiliation{Physics Division, Lawrence Berkeley National Laboratory, 1 Cyclotron Rd, Berkeley, CA 94720, USA}
\affiliation{Berkeley Center for Cosmological Physics, University of California, Berkeley, CA 94720, USA}

\date{\today}

\begin{abstract}
        Noise maps from CMB experiments are generally statistically anisotropic, due to scanning strategies, atmospheric conditions, or instrumental effects. Any mis-modeling of this complex noise can bias the reconstruction of the lensing potential and the measurement of the lensing power spectrum from the observed CMB maps.
        We introduce a new CMB lensing estimator based on the maximum a posteriori (MAP) reconstruction that is minimally sensitive to these instrumental noise biases. By modifying the likelihood to rely exclusively on correlations between CMB map splits with independent noise realizations, we minimize auto-correlations that contribute to biases. 
        In the regime of many independent splits, this maximum closely approximates the optimal MAP reconstruction of the lensing potential.
        In simulations, we demonstrate that this method is able to determine lensing observables that are immune to any noise mis-modeling with a negligible cost in signal-to-noise ratio.
        Our estimator enables unbiased and nearly optimal lensing reconstruction for next-generation CMB surveys. 
\end{abstract}

\maketitle

\section{Introduction}

The gravitational lensing of the cosmic microwave background (CMB) is a powerful probe of the large-scale structure of the Universe. It allows us to reconstruct the projected mass distribution along the line of sight and to constrain the clustering of matter, the sum of neutrino masses and the properties of dark energy~\cite{Lewis:2006fu}.

Quadratic estimators (QEs), which optimally combine weighted pairs of CMB modes, are typically used to reconstruct the lensing potential from the statistical anisotropies it creates in the CMB~\cite{Okamoto:2003zw}. However, it has been established that the QE-approach is suboptimal for low-noise CMB observations, in particular with deep polarization surveys \cite[e.g.,][]{Hirata:2003ka,Carron:2017mqf,Millea:2021had}. While the QE is equivalent to a Gaussian approximation of the lensing potential likelihood, maximum a posteriori (MAP) estimators use all of the information in the CMB maps to reconstruct the lensing potential. For a CMB-S4 configuration, the MAP estimator is expected to halve the errors on the reconstruction noise power spectrum over a wide range of scales, as compared to the QE~\cite{Legrand:2021qdu}.

CMB lensing estimators are sensitive to the modelling of the noise and other sources of statistical anisotropies in the CMB maps. This shows up in two ways.
Firstly, any non-lensing source of anisotropies in the CMB maps, such as masking, scanning strategies and atmospheric effects, can lead to a bias in the reconstructed lensing potential map, known as the \textit{mean-field}. 
Secondly, for the QE, the power spectrum of the reconstructed lensing map is a four-point function of the CMB maps, so it includes the lensing signal we want to measure, plus other four-point contractions.
The dominant term is the disconnected four-point function of the Gaussian CMB and noise (giving rise to what is referred to as the \nlzero bias), which is present even in the absence of gravitational lensing. At small scales, this bias can be orders of magnitude larger than the signal. 
Thus, in order to avoid biases in the lensing map and in the lensing power spectrum, the mean field and \nlzero bias are estimated with Monte Carlo simulations, reproducing the noise pattern and scanning strategy of the CMB experiment \cite{Regan:2010cn, Namikawa:2012pe, Planck:2018lbu, Carron:2022eyg}, and subtracted.

However, accurate simulations of the observed CMB are difficult to achieve, in particular for ground-based surveys with complex atmospheric effects \cite{ACT:2023dou}.
One solution introduced in \cite{Madhavacheril:2020ido} is to take advantage of the fact that the atmospheric and instrumental noise in the CMB maps vary with short correlation times. This means that if we are able to split the CMB maps into several sets observed at different times, they will have independent atmospheric and instrumental noise, while the CMB component (and potential Galactic and extragalactic foregrounds) will be stable.
This \textit{cross-only} QE uses different pairs of CMB maps in each leg of the estimator, and combines all possible pairs avoiding auto-correlations in the lensing power spectrum.
Because the noise in each leg of the QE lensing power spectrum is independent, this estimator effectively nulls the noise component in the mean field and in the \nlzero biases.
Reference~\cite{Madhavacheril:2020ido} demonstrated that in the limit of a very large number of splits, or when the CMB noise level is much lower than the signal, the lensing power spectrum can be reconstructed with a negligible cost in signal-to-noise ratio. 
This allows for an unbiased estimate of the CMB lensing power spectrum, without the need for accurate simulations of the various sources of noise.

Alternatively, \cite{Sherwin:2010ge} proposed a method to remove the disconnected biases completely, i.e., from both noise \emph{and} signal, by applying on the CMB maps mutually exclusive annular filters in  Fourier space. This method was extended  in \cite{Shen:2024vft}, relying on the fact that the Gaussian noise bias comes from a relatively small number of specific multipole configurations of the CMB four-point function, which can be removed from the lensing power spectrum. However, these two methods break for highly anisotropic noise configurations, such as the ones we are interested in our work.

In this paper we extend the cross-only estimator of Ref.~\cite{Madhavacheril:2020ido} to the maximum a posteriori (MAP) lensing estimator. 
The main difficulty is that, by definition, the MAP estimator uses all the information in the CMB maps to reconstruct the lensing potential. So, if we split the observed CMB into several maps observed at different times, the likelihood formalism will leverage both auto- and cross-correlations between these different maps to reconstruct the lensing signal. 
We will develop a new method that allows us to neglect the auto-correlation parts in the likelihood-maximization algorithm. In practice, this comes down to replacing the standard likelihood estimator by a loss function that mimics the likelihood while neglecting the auto-correlations of the split maps. 
While this loss function has in principle no guarantee to converge to the maximum of the likelihood, we show that in scenarios where the CMB signal dominates over the noise, the maximization of the loss function converges to a lensing map very close to optimal, while greatly reducing the noise contribution in the mean field and in the biases of the lensing power spectrum, at a negligible cost in signal-to-noise ratio.

The paper is organised as follows. In Sec.~\ref{sec:formalism} we review the formalism of the MAP lensing estimator and introduce the cross-only MAP estimator. In Sec.~\ref{sec:simulations} we validate the estimator on simulations for different CMB configurations. We conclude in Sec.~\ref{sec:conclusion}. Finally, we derive the mean-field of our cross-only estimator in Appendix~\ref{sec:meanfield}.

\section{Lensing estimators}
\label{sec:formalism}

\subsection{Maximum a posteriori lensing estimator}

We review here the formalism of the maximum a posteriori (MAP) lensing estimator, as introduced in \cite{Carron:2017mqf,Belkner:2023duz}. 
We work here in the flat-sky formalism for simplicity, but the approach naturally extends to the curved sky. 
Note that in Sec.~\ref{sec:simulations}, we will work with full-sky simulations and use curved-sky estimators.

We model the observed CMB maps as the data vector
\begin{equation}
    \Xdat = \Beam\Da \tilde X + n \, .
\end{equation}
Here, the column vector $\Xdat (\hat n) = (T^{\rm dat}, Q^{\rm dat}, U^{\rm dat})^T (\hat n)$ is the observed CMB data vector for temperature and $Q$ and $U$ Stokes parameters in real space; $\tilde X_{\vl} = (T_{\vl}, E_{\vl}, B_{\vl})^T$ are the unlensed CMB fields in multipole space (with $E_{\vl}$ and $B_{\vl}$ the $E$- and $B$-mode polarization multipoles, respectively); \Beam is the beam and transfer function of the instrument; and $n$ the noise in the maps. The lensing deflection field $\va$ is related to the lensing potential with $\va = \nabla\phi$ (we neglect the curl potential), and $\Da$ is the 
map-synthesis and
deflection operator that transforms the unlensed CMB modes 
in multipole space into the lensed Stokes parameters in real space. 

The covariance of this data vector at a fixed lensing deflection field $\va$ is 
\begin{equation}
    \label{eq:cov}
    \Cov_{\va} = \left< \Xdat X^{\rm dat, \dag}\right>_{\va} = \Beam \Da C^{\rm unl} \Da^\dag \Beam^\dag + N \, ,
\end{equation}
where the covariance $C^\text{unl}$ contains the unlensed CMB power spectra and $N$ is the noise covariance matrix. The delensing operation is $\Da^\dag$; an efficient implementation of this operator is discussed in \cite{Reinecke:2023gtp}.

At fixed lensing deflections, the CMB follows a Gaussian likelihood $\mathcal L $. Assuming a Gaussian prior, $-2\mathcal \ln \mathcal P (\va) = \sum_{\vL} \left|\phi_{\vL}\right|^2/C_{\vL}^{\phi \phi}+\text{const.}$, on the lensing field, we can write, up to a constant, the log-posterior of the deflection field as 
\begin{align}
        \ln \mathcal P (\va | \Xdat ) &= \ln \mathcal L (\Xdat | \va)  
        + \ln \mathcal P(\va) \nonumber \\
        &= -\frac 12 X^{\rm dat, T} \Cov_{\va}^{-1} \Xdat - \frac 12 \ln \det \Cov_{\va} \nonumber \\ 
        &\hspace{0.15\textwidth}- \frac 12 \sum_{\vL} \frac{\left|\phi_{\vL}\right|^2}{C_{\vL}^{\phi\phi}} \, .
    \label{eq:logpost}
\end{align}
Following the procedure of \cite{Carron:2017mqf,Belkner:2023duz}, the maximum a posteriori deflection field $\va^{\rm MAP}$ is found by Newton--Raphson iterations on this posterior. This involves computing the gradient and the curvature of the log-posterior with respect to the deflection field. The iterations are converged when the gradient of the log-posterior vanishes.

The gradient of the term which is quadratic in the data can be written as\footnote{Formally, the gradient is evaluated with respect to the two Cartesian
components, $\alpha_1$ and $\alpha_2$,
of the deflection field in the flat sky.} 
\begin{equation}
    \label{eq:gqd}
        g^{\rm QD}_{a}  = \left[\bar X \right]^x \left[\Da \nabla_a X^{\rm WF}\right]_x \, ,
\end{equation}
where the index $a$ denotes one of the two Cartesian coordinates on the flat sky, $a \in (1, 2)$,  and the brackets are for summing over the stokes paramaters $x \in (T, Q, U)$, 
and we have introduced the filtered maps
\begin{equation}
    \begin{split}
        \bar X &\equiv \Beam^\dag \Cov_{\va}^{-1} \Xdat \, , \\ 
        X^{\rm WF} &\equiv  C^{\rm unl} \Da^\dag \Beam^\dag \Cov_{\va}^{-1} X^{\rm dat} \, .
    \end{split}
\end{equation}
Note that the Wiener-filtered field $X^{\text{WF}}$ is defined in harmonic space and the notation $\nabla_a X^{\text{WF}}$ is a shorthand for $i l_a X^{\text{WF}}(\vl)$.

The gradient of the log-determinant of the covariance is a mean field term denoted by $\gMF$. Indeed, since the gradient of a log-likelihood always vanishes in the mean at the true parameter values, and since the covariance does not depend on the data, we can write $\gMF = \left< \gQD \right>_{\va} $,  where the average is taken over realizations of the CMB for a fixed deflection field \va. This \gMF corresponds to a \textit{delensing induced} mean field, where the source of anisotropy is the lensing estimate itself \cite{Legrand:2025vzg}.

Once converged, the lensing map $\phi^{\rm MAP}$ is normalized by a Wiener filter, as shown in \cite{Legrand:2021qdu}. The fiducial value of the Wiener filter is 
\begin{equation}
    \mathcal{W}_L= \frac{\cppfid}{\cppfid + 1/R_L} \, ,
\end{equation}
where the response $R_L$ is obtained iteratively from the partially delensed CMB spectra and the delensing efficiency, following the procedure of \cite{Legrand:2021qdu, Legrand:2023jne}.
However, since the fiducial Wiener filter is only an isotropic approximation of the response of $\phi^{\rm MAP}$ to the true lensing signal, we always correct this normalization with Monte Carlo simulations.

\subsection{Cross-only iterative estimator}

Let us now assume we can split the observed CMB map into a set of $n$ maps $X^{\rm dat}_i$, observed at different times. For simplicity, we assume the same noise properties in each split, so that the optimal co-add map is simply the mean
\begin{equation}
    \Xdat = \frac 1n \sum_{i=1}^n X^{\rm dat}_i \, .
\end{equation}

The quadratic term of the log-likelihood in Eq.~\eqref{eq:logpost} is then given by 
\begin{equation}
    \label{eq:xixj}
    X^{\rm dat, T} \Covai \Xdat = \frac{1}{n^2} \sum_{i, j} X^{\rm dat, T}_i \Covai X^{\rm dat}_j \, ,
\end{equation}
and the quadratic gradient becomes 
\begin{equation}
    \label{eq:gqdsplit}
    \gQD_a = \frac{1}{n^2} \sum_{i,j} \left[ \bar X_i \right]^x \left[\Da \nabla_a X^{\rm WF}_j \right]_x \, .
\end{equation}

Our goal now is to perform an iterative lensing reconstruction which neglects the information coming from the auto-correlations of the split maps. Our approach is to remove all the auto-correlation terms from the quadratic gradient in Eq.~\eqref{eq:gqdsplit}.
This gives the following expression for a cross-only quadratic gradient
\begin{equation}
    \label{eq:gqdx}
    \gQDx_{a} = \frac{1}{n(n-1)} \sum_{i \neq j} \left[\bar X_i\right]^x \left[\Da \nabla_a X^{\rm WF}_j\right]_x \, ,
\end{equation}
where we also rescaled the normalisation by $n/(n-1)$ such that its expectation value over realisations of the CMB fields is equivalent to the co-add gradient of Eq.~\ref{eq:gqdsplit}. 
This cross-only gradient is similar to the cross-only QE of \cite{Madhavacheril:2020ido}, adapted to the MAP formalism with the inclusion of the lensing and delensing operators in the filtering. 

In practice, this cross-only gradient is the only modification we make to the standard MAP estimator. Thus, the filtering operations, as well as the Hessian matrix estimate (following the L-BFGS scheme as in \cite{Carron:2017mqf}) used in the iterations, are unchanged with respect to the standard (co-add) MAP estimator.
We reconstruct the cross-only lensing field \pMAPx, by modifying the \texttt{delensalot}\footnote{\url{https://github.com/NextGenCMB/delensalot}} pipeline to rely on the cross-only gradient \eqref{eq:gqdx} during the iterations, instead of the full co-add one \eqref{eq:gqd}. 

Comparing the cross-only gradient in Eq.~\eqref{eq:gqdx} to the full gradient of Eq.~\eqref{eq:gqdsplit}, we see that the information lost by removing the auto-correlations scales as $\mathcal{O}(1/n)$. This means that in the limit of a large number of splits, there should be no significant loss of information in performing the cross-only iterative reconstruction.

The costliest operation, which cannot be avoided, is to compute the $n$ Wiener-filtered maps.
In principle this Wiener-filtering step is the same for all maps, as the deflected covariance $\Cova$ is the same. However, the matrix inversion cannot be computed exactly due to its high dimensionality, and we use the conjugate-gradient inversion method to estimate the $X_i^{\rm WF}$. This inversion method does not store the matrix inverse, only the product $\Cov_{\va}^{-1}X_i^{\rm dat}$, so we cannot repeat the same operation for all the splits, and we have to perform $n$ conjugate-gradient inversions at each iteration.

\subsection{Loss function}

Let us remark that the cross-only gradient in Eq.~\eqref{eq:gqdx} can be obtained from a loss function, which mimics the likelihood of the MAP estimator but neglects the auto-correlations of the split maps. This loss function is given by
\begin{multline}
        \label{eq:lossfunc}
        f(\va, \Xdatx) = -\frac{1}{2} \Xdatx^{\rm T} K_{\va}^{-1} \Xdatx  - \frac 12 \ln \det \Cova \\ 
        - \frac 12 \text{Tr}\left[ \Covai N \right] - \frac 12 \sum_{\vL} \frac{\left|\phi_{\vL}\right|^2}{C_{\vL}^{\phi\phi}} \, ,
\end{multline}
where we introduced the split data vector
\begin{equation}
    \label{eq:splitdat}
    \Xdatx = \left(X^{\rm dat}_1, X^{\rm dat}_2, \dots,  X^{\rm dat}_n\right)^{\rm T} \, ,
\end{equation}
and the block matrix $K_{\va}^{-1}$, consisting of $n \times n$ blocks, which is zero on the block diagonal and $\Cov_{\va}^{-1}$ elsewhere:
\begin{equation}
    \label{eq:K}
    K_{\va}^{-1} = \frac{1}{n(n-1)} \begin{pmatrix}
        0  & \Cov_{\va}^{-1} & \dots & \Cov_{\va}^{-1} \\
        \Cov_{\va}^{-1} & 0 &  \dots & \Cov_{\va}^{-1} \\
        \vdots &  \vdots & \ddots & \vdots \\
        \Cov_{\va}^{-1} & \Cov_{\va}^{-1}  & \dots & 0 
    \end{pmatrix} \, .
\end{equation}

The gradient of the quadratic term of the loss function in Eq.~\eqref{eq:lossfunc} with respect to the deflection field gives back the cross-only gradient in Eq.~\eqref{eq:gqdx}.
The block matrix $K_{\va}^{-1}$ mimics the covariance matrix of the standard likelihood, but is not positive-definite, although it is invertible. So it is not a covariance matrix and the loss function we introduced is not a likelihood anymore. 
This means that this loss function might not have an extremum. 
However, as validated in Sec.~\ref{sec:simulations} with simulations, in the limit of signal-dominated maps, the quadratic form becomes in practice positive-definite.

Indeed, assuming $X_i^{\rm dat} = X^{\rm dat} + n_i$, the quadratic products between the $i$ and $j$ maps, with $i\neq j$, is
\begin{equation}
    \begin{split}
        X_i^{\rm dat, T} \Cov_{\va}^{-1} X_j^{\rm dat} &= X^{\rm dat, T} \Cov_{\va}^{-1} X^{\rm dat} \\ &+ X^{\rm dat, T} \Cov_{\va}^{-1} n_j +  n_i^{\rm T} \Cov_{\va}^{-1} X^{\rm dat}  \\ 
        &+ n_i^{\rm T} \Cov_{\va}^{-1} n_j \, .
    \end{split}
\end{equation}
When the noise is negligible with respect to the data, we see that the quadratic form becomes in practice positive-definite, as only the part $X^{\rm dat, T} \Cov_{\va}^{-1} X^{\rm dat}$ contributes.

The mean-field part -- i.e., the second and third term in Eq.~\eqref{eq:lossfunc} -- is defined such that the loss function is kept unbiased, in the sense that the gradient of the loss function (without the prior) vanishes in the mean at the true lensing potential (see the derivation in Appendix~\ref{sec:meanfield}).
The gradient of the extra term $\text{Tr}\left[ \Covai N \right]$ corresponds to the mean-field sourced by the delensing of the noise, as discussed in detail in Ref.~\cite{Legrand:2025vzg}. This delensed noise mean-field comes from anisotropies created in the noise map when delensing the CMB. 
As can be inferred from Eq.~\eqref{eq:lossfunc}, the cross-only estimator does not contain the noise contribution to the delensing induced mean-field, which is in principle included in the standard MAP mean-field given by the gradient of $\ln \det \Cova$.
In practice we will neglect the delensing induced mean-field in the iterations. In \cite{Legrand:2025vzg} we showed that this mean-field term does not impact the lensing reconstruction cross-correlation coefficient. It appears only to impact the normalisation of the estimated lensing field. So, provided the normalisation is correctly estimated with Monte Carlo simulations, we can safely neglect the delensing induced mean-field.

\subsection{Summary of the iterative process}
\label{subsec:iterative_summary}

We summarize below the iterative procedure of the cross-only MAP:
\begin{enumerate}
    \item Get an estimate of the lensing deflection field $\boldsymbol{\alpha}_N$. For the first step we take the cross-only QE of \cite{Madhavacheril:2020ido}.
    \item Compute the $n$ Wiener-filtered delensed split maps using a conjugate-gradient inversion.
    \item Estimate the split gradient $\boldsymbol{g}^{\rm QD, \times}$ from Eq.~\eqref{eq:gqdx}.
    \item Add the prior gradient and obtain the total gradient $\boldsymbol{g}_N$ (we neglect the delensed mean-field gradient).
    \item The Hessian curvature matrix $H_N$ (i.e., the second derivative of the log-posterior) is updated following the L-BFGS scheme as in \cite{Carron:2017mqf}.
    \item We obtain the next lensing deflection field as $\va_{N+1} = \va_N + \lambda H_N \boldsymbol{g}_N $, where we take $\lambda=0.5$ to improve convergence. 
\end{enumerate}

We denote by \pMAPx the cross-only lensing potential obtained at convergence. Its effective normalization is determined using a set of Monte Carlo simulations, following the same procedure as for the standard MAP estimator.

Finally, we can estimate the mean-field coming from mask, noise or other anisotropic features not due to lensing, by averaging \pMAPx on a set of realistic Monte Carlo simulations, such as in \cite{Legrand:2023jne}. This mean-field is then subtracted from the estimated lensing field.

In our iterative delensing procedure, we use the deflection field estimated from split maps, $\va_N$, to delens all the maps at the next iteration. This approach does not eliminate all noise contractions in the quadratic gradient estimator: higher-order contractions involving the same maps can remain.

To illustrate this schematically at the first iteration, consider two split maps, $X_1$ and $X_2$. The cross-only quadratic estimator (QE) for the lensing deflection can be written as in Eq.~\ref{eq:gqdx} (for $\va = 0$)\footnote{There are 
implicit $\mathcal{D}_{\va = 0}$ map-synthesis operators in the Wiener-filtered fields here, which we suppress to avoid clutter.}:
\begin{equation}
    \va^{12}_{(0)} \sim \frac{1}{2} \left( \bar X_1 \nabla X_2^{\rm WF} + \bar X_2 \nabla X_1^{\rm WF} \right) \, .
\end{equation}
The delensed CMB map at the first iteration is then approximately
\begin{equation}
X^{\rm del}_1 \simeq X_1 - \va^{12}_{(0)} \cdot \nabla X_1 \, .
\end{equation}
Assuming the MAP estimate behaves like a QE applied to delensed maps, the first iteration of the deflection field will contain contractions of the form 
\begin{equation}
        \label{eq:firstiter}
        \va^{12}_{(1)} \sim (X_1 - \va^{12}_{(0)}\cdot \nabla X_1) (X_2 - \va^{12}_{(0)}\cdot \nabla X_2 ) \, .
\end{equation}
This expression contains residual contractions such as $X_1 X_1$ and $X_2 X_2$ through $\va^{12}_{(0)}$,  
implying that the noise is not completely cancelled in the iterative process. In this context, this corresponds to the noise contribution to the mean-field.
In principle, one could devise a more optimal scheme in which pairs of maps $X_i$ and $X_j$ are delensed using a deflection field $\va_{N}$ that is independent of both maps $i$ and $j$. However, we find that our simpler method already suppresses most of the mean field noise contractions effectively, as demonstrated in Sec.~\ref{sec:sims_meanfield}.

\begin{figure*}
    \includegraphics[width=0.9\textwidth]{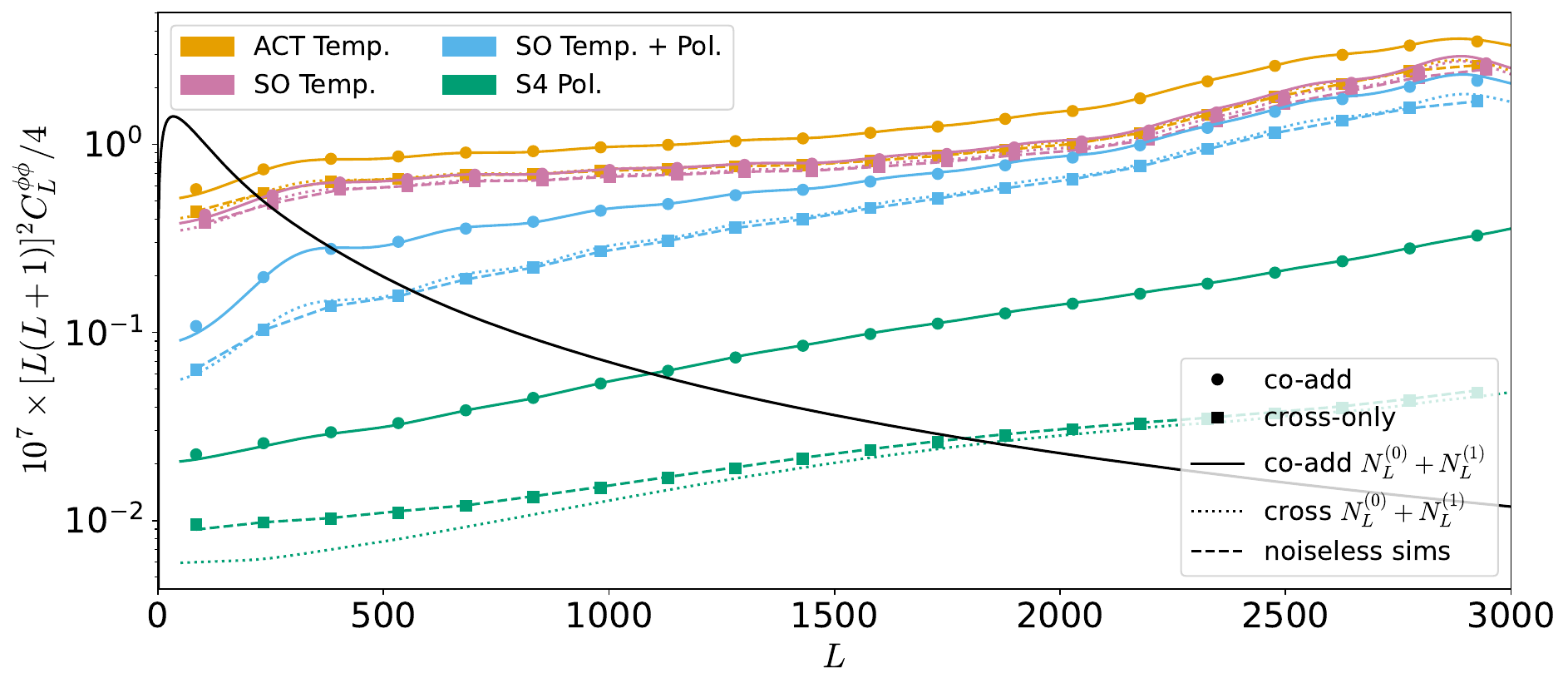}
    \caption{Biases of the MAP and cross-only MAP estimators for our three experimental configurations. The mean biases obtained from 200 simulations with the MAP and cross-only MAP estimators are shown as the circles and squares, respectively. The solid and dotted lines are the predictions of the \nlzero and \nlone biases for the MAP and the cross-only MAP estimators, respectively. The dashed lines are the residual biases obtained from 100 noiseless simulations with the standard (co-add) MAP estimator for each experimental configuration. The black solid line is the fiducial CMB lensing power spectrum.
    We show ACT-like temperature (orange), SO-like for temperature and for both temperature and polarization (purple and sky blue) and CMB-S4-like polarization (teal). We see that the CMB-S4 polarization configuration benefits the most from the bias reduction from the split estimator, while when the CMB fields are signal dominated, such as for the SO temperature case, the bias is less significantly reduced. Note that this bias reduction does not correspond to a reduction in the variance of the band-power estimates. 
    }
    \label{fig:noise_bias}
\end{figure*}

\subsection{Power spectrum from splits}
\label{sec:powerspectrum}

The cross-only QE power spectrum, as introduced in \cite{Madhavacheril:2020ido}, is defined as the sum of the lensing field estimated from all possible pairs of maps, avoiding any auto-correlation:
\begin{equation}
    \label{eq:crossqe_power}
    C_L^{\rm QE, \times} = \frac{1}{n(n-1)(n-2)(n-3)}\sum_{i\neq j \neq k \neq l} C_L(\hat \phi^{\rm QE}_{ij}, \hat \phi^{\rm QE}_{kl}) \, ,
\end{equation}
where $\hat \phi^{\rm QE}_{ij}$ means a QE lensing field estimated from the pair of maps $i$ and $j$.

However, for the cross-only MAP we introduced in Sec.~\ref{sec:formalism}, even if the gradient \gQDx in Eq.~\eqref{eq:gqdx} can be decomposed into a sum of quadratic terms, the \pMAPx obtained at convergence cannot be separated into a sum of quadratic terms with pairs of CMB maps. Indeed, at each iteration we apply the lensing and delensing operators, based on the current estimate of \pMAPx, so the final lensing map is a complex high-order combination of the split maps.

Assuming we can decompose the CMB maps into $n$ splits, it is possible to estimate \pMAPx from any subset of these $n$ splits. For example, we can take a subset $a$ that contains $k$ maps to estimate $\pMAPx_a$, and the complementary subset $b$ that contains the $n-k$ other maps to estimate $\pMAPx_b$. The power spectrum $C_L(\pMAPx_a, \pMAPx_b)$ will avoid repetitions in the pairs of maps. 

It follows that, if we have $n$ splits, we can estimate the lensing power spectrum with any combinations of the two subsets of maps of size $k$ and $n-k$, for all $k$. 
\begin{equation}
    C_L^{\rm MAP, \times} = \frac{1}{N_{\rm sets}}\sum_{a, b} C_L(\pMAPx_a, \pMAPx_b) \, ,
\end{equation}
where $N_{\rm sets}$ is the number of sets to average over.
The number of subsets scales as $\mathcal{O}(2^n)$, which becomes quickly infeasible for large $n$. 
In a simple and more realistic scenario where $n=4$, there are only six possible subsets, each involving a pair of maps.
The cross-only MAP power spectrum is then simply
\begin{multline}
    \label{eq:cppmapx}
        C_L^{\rm MAP, \times} = \frac{1}{3} \left[ C_L(\pMAPx_{12}, \pMAPx_{34}) \right. \\ 
        + C_L(\pMAPx_{13}, \pMAPx_{24}) + \left. C_L(\pMAPx_{14}, \pMAPx_{23}) \right] \, ,
\end{multline}
which is the same form as for the cross-only QE power spectrum.
 
For the standard quadratic estimator, the estimated lensing spectra contain the signal we want to measure plus biases due to other contractions in the 4-point function of the CMB. Schematically, we have 
\begin{equation}
    C_L^{\hat \phi \hat \phi} = \hat C_L^{\phi \phi} + \nlzero + \nlone \, ,
\end{equation}
with \nlzero the disconnected 4-point function and \nlone the connected 4-point function that is first order in \cpp.

The power spectrum of the MAP also contains biases, noted \nlzeroMAP and \nloneMAP by analogy with the QE. In principle these are not only 4-point functions of the CMB maps anymore, as evident from the structure of the first iteration in Eq.~\ref{eq:firstiter}. However we can predict theses biases using partially delensed CMB spectra in the standard expression of the QE biases as shown in \cite{Legrand:2021qdu}.

For the cross-only MAP power spectrum, the disconnected contractions in the cross-power spectrum $C_L(\pMAPx_{12}, \pMAPx_{34})$ that contributes to the $N_L^{(0)}$ bias, such as $X_1X_3$ and $X_2X_4$, are, by construction, free of noise contributions, since they avoid any auto-correlations between the split maps.
In Section~\ref{sec:sims_isotropic} we will confirm with noiseless simulations that the cross-only MAP $N_L^{(0), \times}$ bias is indeed free from noise contributions.

We predict analytically the cross-only MAP $N_L^{(0), \times}$ bias by setting to zero the noise spectra in the standard  \nlzeroMAP bias procedure.
The \nlone bias is assumed to be independent of the noise in the CMB maps, as is the case for the QE estimator. So we keep the same expression for the standard MAP and the cross-only MAP.
The sum of these biases are shown in Fig.~\ref{fig:noise_bias} as the solid and dotted lines for the standard and cross-only MAP estimators, respectively.

\section{Validation with simulations}
\label{sec:simulations}

We now test our cross estimator and demonstrate that it converges to a close-to-optimal lensing field while being able to reduce both the lensing reconstruction noise at high $L$ and the mean-field bias at low $L$.

\begin{table}
    \centering
    \begin{tabular}{c c c c c} \hline \hline
         & Noise level $\Delta T$ & Beam & $\ell_{\rm min}$ & $\ell_{\rm max}$ \\
         & [$\mu\text{K-arcmin}]$] & [arcmin] & & \\
        \hline
        ACT & 10 & 3 & 40 & 3000  \\
        SO & 6 & 1.4 & 100 & 3000 \\
        CMB-S4 & 1 & 1 & 40 & 4000 \\ \hline
    \end{tabular}
    \caption{Experimental configurations for the simulations. We consider configurations similar to ACT, Simons Observatory (SO) and CMB-S4. The polarisation noise levels are given by $\Delta P = \sqrt{2} \Delta T$. The  $\ell_{\rm min}$ and  $\ell_{\rm max}$ are the CMB multipole range used for estimating the lensing potential field.}
    \label{tab:exp}
\end{table}

\subsection{Isotropic noise}
\label{sec:sims_isotropic}

We simulate CMB temperature and polarization full-sky maps, lensed by a Gaussian lensing potential field, and add Gaussian noise realizations. We consider three different experimental configurations, mimicking current and upcoming surveys, as listed in the Table \ref{tab:exp}. We make 200 simulations of the CMB for each experimental configuration.
To mimic the split of the data into four maps, for each simulation we generate four Gaussian noise fields, with the noise levels increased by a factor of two in each map. 

We reconstruct the lensing potential using the temperature maps for the ACT-like configuration, either the temperature only, or both temperature and polarization maps, for the SO-like configuration, and only the polarization maps for the CMB-S4-like configuration.
For each of the 200 simulations, and each experiment, we reconstruct lensing with the QE and MAP estimator and with the cross-only QE and cross-only MAP estimator. For the MAP estimators, we run five iterations on each map, as we found that these achieve sufficient convergence. We estimate the effective normalization of \pMAP and \pMAPx using the average of the cross-correlation of the 200 reconstructed maps with the true lensing fields, as in \cite{Legrand:2021qdu}.

We show in Fig.~\ref{fig:noise_bias} the biases on the lensing power spectrum for the MAP and cross-only MAP estimators, averaged over the 200 simulations. 
The circles and squares are the binned average of the lensing power spectra from \pMAP and \pMAPx, where we subtracted the input lensing spectrum. The predictions for the MAP biases, $\nlzero + \nlone$, are shown with solid lines, and the cross-only MAP biases, $N_L^{0, \times} + N_L^{1}$, are shown as the dotted lines.

The noise  bias is reduced by a factor of around five for the CMB-S4 polarization case. The reduction is less significant in the SO temperature case. This is expected as the CMB temperature is totally signal dominated in the range of multipoles we considered for the SO configuration (up to $\ell =3000$), so the bias reduction from cancellation of the noise auto-correlations is negligible.
We see that there is a good agreement between the predicted and estimated noise biases for both the MAP and cross-only MAP estimators. The prediction for cross-only MAP with the CMB-S4 configuration is around $10\%$ lower than the simulation results for $L\sim 1500$, and around $50\%$ lower for $L\sim 200$. 
In a realistic analysis, this small offset would be captured by a realisation-dependent bias subtraction (\nlrdzero) \cite{Hanson:2010}, and by a Monte Carlo correction applied on the lensing bandpowers, but neither are applied here.

To directly isolate the impact of noise, we generate 100 noiseless CMB simulations and reconstruct the lensing potential using the standard MAP estimator, applying the fiducial noise level in the filtering corresponding to each experimental setup. 
The residual biases from these reconstructions, plotted as the dashed lines in Fig.~\ref{fig:noise_bias}, match precisely the noise biases observed in our cross-only MAP reconstructions across all configurations.
This agreement provides compelling evidence that our cross-only MAP estimator entirely cancels the noise contribution to the lensing power spectrum biases.

\begin{figure}
    \includegraphics[width=\columnwidth]{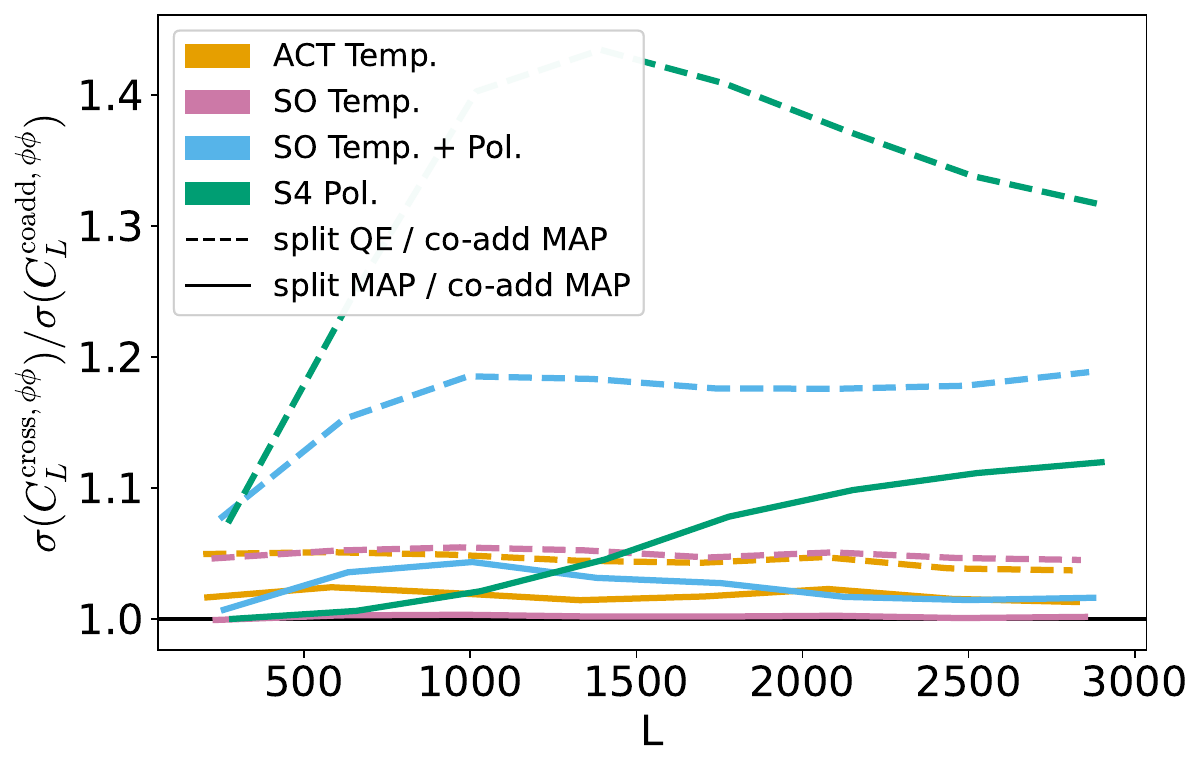}
    \caption{Ratio of the CMB lensing power spectrum variance, for the cross-only estimators over the standard (co-add) MAP estimator, obtained from 200 full-sky simulations and binned in bandpowers. We show the cross-only QE (dashed lines) and cross-only MAP (solid lines), over the co-add MAP, for our three experimental configurations. For the SO-like case we show both the temperature-only and the temperature and polarization estimators. The variance of the cross-only estimators are always higher than the co-add MAP estimator, as expected. We see that the increase in variance of the cross-only QE is greatly reduced for the cross-only MAP estimator, in particular for surveys including polarization. Comparing the cross-only MAP to the co-add MAP, we see that the increase in variance is below $5\%$ for the ACT-like and SO-like configurations.
    For the CMB-S4 case, at multipoles $L<1000$, where most of the signal-to-noise is accumulated, the degradation is still negligible.}
    \label{fig:sigma}
\end{figure}

Even if the biases are reduced with the cross-only MAP estimator, this does not correspond to a reduction in the variance of the lensing bandpowers over our set of simulations. 
We compare in Fig.~\ref{fig:sigma} the bandpower errors for the cross-only estimators over the standard co-add MAP estimator. We see that the MAP cross-only estimator has a lower increase in variance than the cross-only QE estimator, in particular for the CMB-S4 polarization and the SO minimum-variance cases. But even for the ACT-like temperature configuration, the increase is reduced by a factor two compared to that for the cross-only QE.
With the cross-only MAP in the temperature reconstruction, the increase in errors compared to the co-add MAP is negligible. For the minimum-variance SO-like configuration, the increase in variance of the cross-only MAP bandowers compared to the co-add MAP ones is below $5\%$ for all multipoles.
In the CMB-S4-like polarization case, the errors for the cross-only MAP estimator are around 10\% larger than the co-adds at high $L$, but below $L=1000$ (which in practice achieves most of the signal-to-noise ratio) the increase is below $5\%$ and so negligible.

\begin{figure}
    \centering
    \includegraphics[width=\columnwidth]{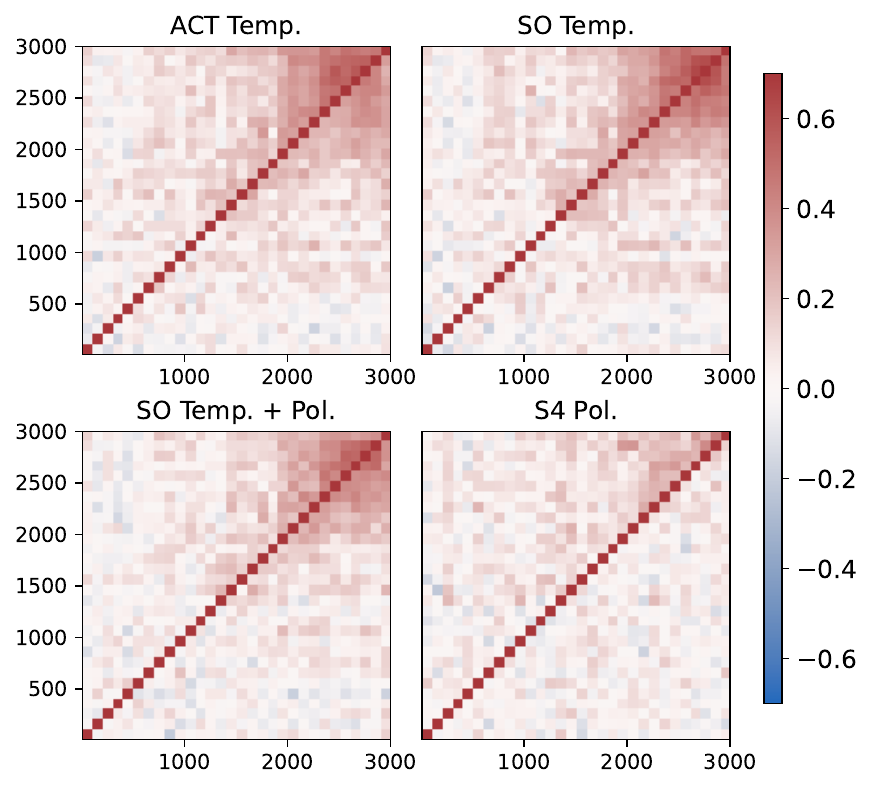}
    \caption{Correlation matrices for the lensing power spectrum \cpp in 25 bins between $L=2$ and $L=3000$. We show the standard MAP in the upper-left triangles, and the cross-only MAP in the lower-right triangles, for the four experimental configurations considered. We see that in the CMB-S4 polarization configuration, the cross-only covariance is visibly more diagonal than the standard MAP. Note that no realisation-dependent bias correction is made, which would significantly reduce the off-diagonal correlations.}
    \label{fig:correl}
\end{figure}

Fig.~\ref{fig:correl} shows the power spectrum correlation matrices, i.e., the covariance matrices normalized by their diagonals, for the MAP and cross-only MAP estimators. These are estimated from our 200 full-sky isotropic-noise simulations. Upper-left triangles are the standard MAP and the lower-right are the cross-only MAP estimators. We see that the correlations from the cross-only estimator are similar to those of the standard MAP estimator, and even slightly more diagonal for the CMB-S4 polarization configuration.
By analogy with the QE, this could come from the reduction of the two-point function for some of the four products of two-point contractions that enter the dominant contribution to the off-diagonal covariance. The relevant terms are given in Eq.~(46c) of~\cite{Hanson:2010}. The gain from not having noise in some of these is larger for polarization than temperature since the latter is almost signal-dominated.

Note that here we do not subtract realisation-dependent forms of the bias corrections. In practice this would greatly reduce the non-diagonal correlations~\cite{Hanson:2010, Legrand:2021qdu}. 

\subsection{Anisotropic noise and mean-field}
\label{sec:sims_meanfield}

Anisotropies in the CMB maps not due to lensing, such as anisotropic noise patterns or the mask of the survey, can bias the reconstructed lensing field. This contribution, called the mean-field, is in general estimated from survey-specific Monte Carlo simulations, and subtracted from the measured lensing map. This requires the simulations to capture the survey strategy and atmospheric noise correlations accurately. In this section, we investigate the mean-field due to time-varying noise (such as instrumental or atmospheric noise), which would create such anisotropies in the CMB maps. Since this noise is time-varying, the cross-only QE built from map splits with independent noise realisations should be able to cancel exactly the noise contribution to the mean-field~\cite{Madhavacheril:2020ido}.

\begin{figure}
    \centering
    \includegraphics[width=\columnwidth]{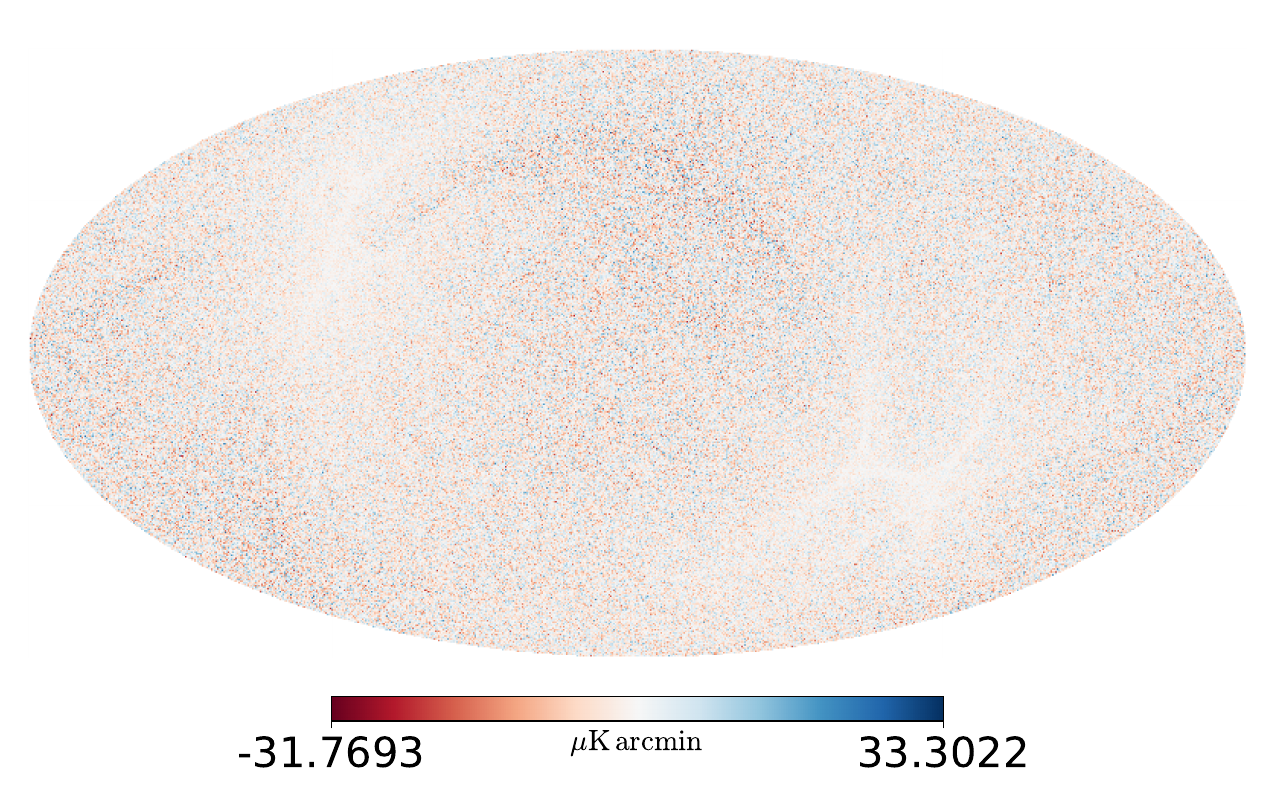}
    \caption{One realization of the noise map used for the simulations, following the Planck scanning strategy, scaled such that the its standard deviation is of $10\,\mu\text{K-arcmin}$. 
    }
    \label{fig:noisemap}
\end{figure}

We generate maps with highly anisotropic noise variance based on the Planck simulations of noise in the 100\,GHz channel. An example noise map is shown in Fig.~\ref{fig:noisemap}. We rescale this map such that its variance is equal to the Gaussian noise level of the simulation settings we consider.  
For each CMB simulation, we generate four map splits by adding independent realisations of the anisotropic noise.

The mean field is estimated from a set of 100 simulations, where we reconstruct the QE, the MAP and the cross-only MAP lensing fields. 
Since we want to test the non-ideal case where we do not know the noise maps, we assume isotropic noise in the filtering operations of the QE and MAP estimators.
To eliminate bias due to common simulations in each averaged mean-field estimate, the mean-field power spectrum is estimated from averaging two sets of 50 reconstructed maps, and taking the cross correlation between these two sets of maps:
\begin{equation}
    C_L^{\phi^{\rm MF}} = C_L\left(\phi^{\rm MF 1}, \phi^{\rm MF 2}\right) \, ,  
\end{equation}
with $\phi^{\rm MF 1}$ the mean-field obtained from averaging the reconstructed lensing field from the first set of simulations and $\phi^{\rm MF 2}$ from the second set. 

For the cross-only estimator we consider four split maps, which gives six pairs of CMB maps to estimate the lensing field from. We estimate the lensing mean-field by averaging the cross-only MAP reconstructed for each pair of maps. We estimate the mean-field power spectrum following the expression of Eq.~\eqref{eq:cppmapx} and computing the cross-spectra with a different set of simulations on each leg to avoid biases: 
\begin{equation}
        C_L^{\rm MF ,\times} = \frac{1}{36} \sum_{i\neq j, k \neq l} C_L\left(\langle\pMAPx_{ij}\rangle_{a}, \langle\pMAPx_{kl}\rangle_{b}\right) \, .
\end{equation}
where the $\langle \, \rangle_{a}$ and $\langle \, \rangle_{b}$ are for averaging over a different subset, each with 50 simulations.

\begin{figure}
    \centering
    \includegraphics[width=\columnwidth]{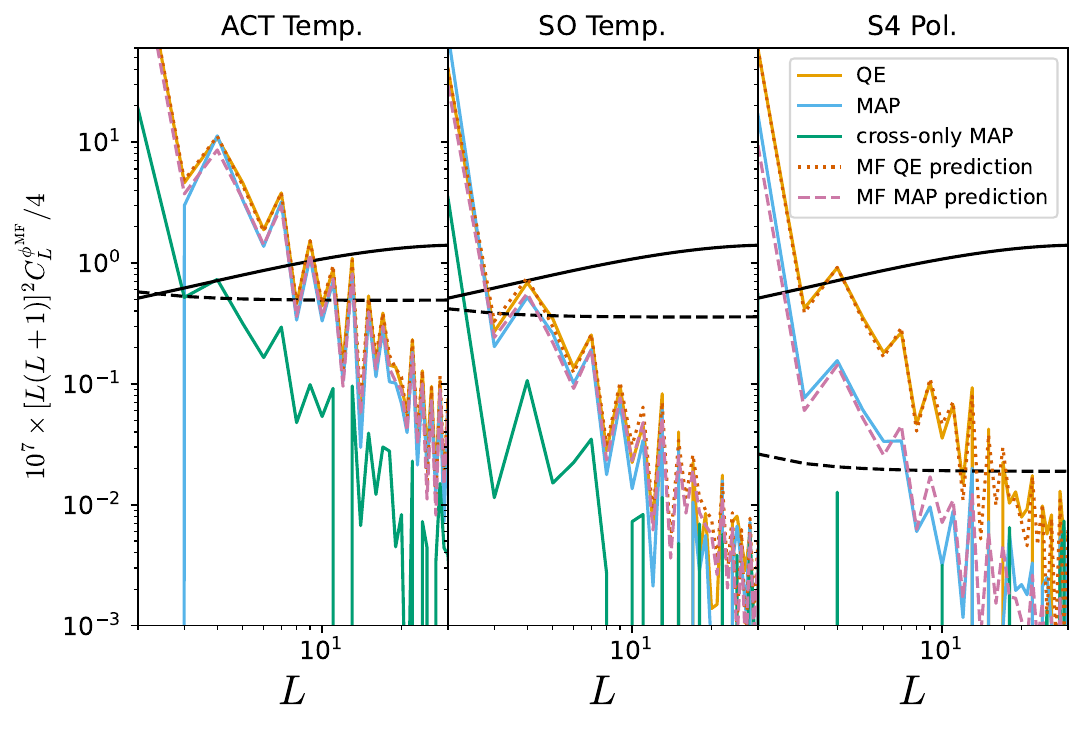}
    \caption{Power spectra of the mean-field for the QE (orange), standard MAP (blue) and cross-only MAP (teal), estimated from averaging over 100 simulations with anisotropic noise patterns. The QE cross-only mean-field is zero by construction. For reference we show the fiducial lensing power spectrum (solid black line) and the \nlzeroMAP bias (dashed black line). The left panel is for the ACT-like configuration with temperature, the central panel is for a SO-like configuration with temperature, and the right panel is for CMB-S4 with polarization. We see that the cross-only MAP is able to reduce the mean-field term, in particular in the polarization CMB-S4 configuration. The mean-field does not disappear in the temperature only reconstructions, but it is reduced by an order of magnitude for both ACT and SO. We also show the predicted mean-field power spectrum due to noise anisotropies as the dotted and dashed lines, for  standard QE and MAP estimators, respectively.}
    \label{fig:meanfield}
\end{figure}

In Fig.~\ref{fig:meanfield} we show the mean-field power spectrum for the QE, the standard MAP and the cross-only MAP. The cross-only QE mean field is zero by construction since we are considering full-sky observations.
The right panel shows the mean-field for the CMB-S4-like polarization configurations. The MAP mean-field is already an order of magnitude lower than the QE mean-field, and we see that cross-only MAP mean-field is completely negligible. 

In the left and right panels of Fig.~\ref{fig:meanfield} we show the mean-field for the ACT-like and SO-like temperature configurations. The QE and MAP mean-fields are very similar since in these configurations the iterative estimator is almost equivalent to the QE. However, we see that the cross-only MAP reduces the mean-field power by about an order of magnitude in both cases. This mean-field is not totally nulled, contrary to the cross-only QE. This is likely due to the fact that during the iterative process, there are some non-trivial combinations of maps involved, as described schematically in Eq.~\ref{eq:firstiter}.

We show as dotted lines the predicted mean-field power spectra arising from noise anisotropies for both the standard QE and MAP estimators. These predictions are obtained using the mean-field (MF) response derived in Appendix B of \cite{Planck:2018lbu}, generalized here to include the polarization-based estimator. For the MAP estimator, the MF response is computed using delensed CMB spectra (obtained via the iterative procedure employed in the calculation of the \nlzero bias) in place of the lensed spectra. These predictions are found to be in good agreement with the mean-field power spectra estimated from simulations.

\section{Conclusion}
\label{sec:conclusion}

In this work we modified the standard maximum a posteriori lensing estimator to allow for cross-only lensing reconstruction from CMB maps with independent noise.
This allows removal of most of the contributions from the auto-correlations of the noise in the CMB maps. We showed that, despite the lack of formal guarantee of convergence, it is possible to maximize the loss-function we introduced and reconstruct a lensing field that is close to optimal. This is especially true when the signal dominates the noise, as in our SO-like temperature-only configuration. 

As for the cross-only QE, our cross-only MAP estimator is able to reduce the noise biases, as well as the mean-field bias coming from noise. The reduction of these biases is most important for polarization-based lensing reconstruction, such as our CMB-S4-like polarization-only configuration, since the noise is large on the small scales of interest for the lensing. The cross-only estimator also reduces the off-diagonal correlations between the CMB lensing band-powers.

We did not investigate the implementation of
realization-dependent bias subtraction (\nlrdzero; see~\cite{Planck:2018lbu} and references therein). This method provides a first-order accurate estimate of the \nlzero bias in the presence of unmodelled survey anisotropies, and reduces correlations between multipoles in the debiased CMB lensing spectra. 
In principle, one could start with the formalism of the cross-only QE \nlrdzero introduced in \cite{Madhavacheril:2020ido}, and extend it to the \nlrdmapzero framework of \cite{Legrand:2021qdu,Legrand:2023jne}, thereby deriving a realization-dependent debiaser for our cross-only MAP estimator. 
However, we leave this extension for future work. Importantly, we have shown that the covariance of the cross-only MAP is slightly more diagonal than that of the standard MAP, and the cross-only \nlzero bias is lower. This suggests that the impact of realization-dependent bias corrections may be less significant in our case.

The cross-only estimator we introduced is close to the optimal lensing reconstruction, and robust to the complex noise properties that arise from ground-based observations.
It will be a useful estimator to validate a standard MAP lensing analysis. Indeed, comparing the reconstruction between different splits of the data, such as in \cite{Planck:2018lbu}, is important to check for potential unknown systematics. 

Finally, our formalism opens the door to new extensions of the QE tools into the MAP formalism. For instance, following the gradient-leg cleaning estimator from \cite{Madhavacheril:2018bxi, Darwish:2020fwf}, we should be able to define an iterative estimator where one leg (the one with the Wiener filtering) of the quadratic gradient has been foreground cleaned, while the other leg retains the full information. One could also extend the halo lensing estimator of \cite{Hu:2007bt}, with one leg with a lower $\ell_{\text{max}}$ filtering.

Our work suggests that the MAP formalism is robust and sufficiently flexible that the tools developed for the QE can be extended to the MAP \cite[see also][]{Carron:2025wqb, Darwish:2025fip}. This is a promising result for the next generation of CMB surveys, where the MAP formalism will be necessary to extract the maximum information from the data.

\section*{Acknowledgments}

The authors thank Niall MacCrann and Jack Kwok for helpful discussions during the early stages of this work.
LL acknowledges support from the Swiss National Science Foundation through the Postdoc.Mobility fellowship (Grant No. P500PT\_21795).
BDS acknowledges support from the European Research Council (ERC) under the European Union’s Horizon 2020 research and innovation programme (Grant agreement No. 851274).
AC acknowledges support from the STFC (grant numbers ST/W000977/1 and ST/X006387/1).
JC acknowledges support from a SNSF Eccellenza Professorial Fellowship (No. 186879).
GSF is supported by Lawrence Berkeley National Laboratory and the Director, Office of Science, Office of High Energy Physics of the U.S. Department of Energy under Contract No. DE-AC02-05CH11231.
Part of this work uses resources provided by the Cambridge Service for Data Driven Discovery (CSD3) operated by the University of Cambridge Research Computing Service (www.csd3.cam.ac.uk), provided by Dell EMC and Intel using Tier-2 funding from the Engineering and Physical Sciences Research Council (capital grant EP/T022159/1), and DiRAC funding from the Science and Technology Facilities Council (www.dirac.ac.uk).

\appendix

 \section{Mean-field of the cross-only estimator}
\label{sec:meanfield} 

 We derive here the expression for the additional mean-field term, $-\tr [\Cova^{-1} N ]/2$, that we introduce in the loss-function~\eqref{eq:lossfunc} for our split-only estimator. 
Assuming we split the data into $n$ split maps, we rewrite the block matrix $K_{\va}^{-1}$ from Eq.~\eqref{eq:K} as a Kronecker product
 \begin{equation}
     K_{\va}^{-1} = \frac{1}{n(n-1)}(J_n - I_n) \otimes \Cova^{-1} \, .
 \end{equation}
Here, $J_n$ is the square matrix of size $n$ with all elements equal to unity and $I_n$ is the identity matrix of size $n$.
 The covariance between two split maps $X_i$ and $X_j$ is
 \begin{align}
     \Cov(X_i, X_j) &= \mathcal{B}\Da C^{\rm unl} \Da^\dagger \mathcal{B}^\dagger + \delta_{ij} N_\times \nonumber \\
     &= \Cova - N + \delta_{ij} N_\times \, ,
 \end{align}
 with $N_\times$ the noise covariance of the split maps and $\delta_{ij}$ the Kronecker delta.
 The main assumption is that the noise is uncorrelated between splits. For simplicity we also assume that the splits receive an equal noise contribution given by $N_\times$, so that $N_\times = n N$, but that assumption can be relaxed.
 The covariance of the split data vector \Xdatx from Eq.~\eqref{eq:splitdat} is then
 \begin{equation}
     \Cov(\Xdatx) = J_n \otimes (\Cova - N) + I_n \otimes N_\times \, .
 \end{equation}

We want the gradient of the loss function (minus the prior term) to vanish at any $\va$ when averaged over data lensed by that same $\va$. More generally, let us distinguish between a point in parameter space, $\va$, and the actual lensing of the data, which we denote with a tilde.
 Averaging the quadratic part of the loss function defined in Eq.~\eqref{eq:lossfunc} over CMB realizations with fixed lensing field $\tilde{\va}$ we get 
 \begin{widetext}
 
 \begin{equation}
    \label{eq:meanfield}
     \begin{split}
         \left<-\frac{1}{2} \Xdatx^T K_{\times}^{-1} \Xdatx\right>_{\tilde{\va}} &= -\frac12 \tr\left[ K_\times^{-1} \widetilde{\Cov}(\Xdatx)\right] \\
         &= -\frac12 \tr\left[ \left( \frac{1}{n(n-1)}(J_n - I_n) \otimes \Cova^{-1} \right) \left( J_n \otimes (\widetilde{\Cova} - N) + I_n \otimes N_\times \right) \right] \\
         &= -\frac12 \frac{1}{n(n-1)} \tr \left[ (J_n - I_n)J_n \otimes \Cova^{-1} (\widetilde{\Cova} - N) + (J_n- I_n)\otimes\Cova^{-1}  N_\times \right] \\  
         &= -\frac12 \frac{1}{n(n-1)} \tr \left[ (n-1) J_n \otimes (\Cova^{-1}\widetilde{\Cova} - \Cova^{-1} N )\right] \\
         &= -\frac12 \tr \left[ \Cova^{-1}\widetilde{\Cova} \right]  +\frac12 \tr \left[\Cova^{-1}N \right] \, .
     \end{split}
 \end{equation}
 \end{widetext}

 Taking the gradient of this expectation with respect to $\va$, and then setting $\tilde{\va} = \va$, is equivalent to taking the gradient of $-\Xdatx^T K_{\times}^{-1} \Xdatx/2$ first and then averaging over data lensed by fixed $\va$. Applied to the right of Eq.~\eqref{eq:meanfield}, this gives
\begin{multline}
-\frac12 \tr \left[ (\nabla_{\va} \Cova^{-1})\Cova \right]  +\frac12 \tr \left[(\nabla_{\va} \Cova^{-1})N \right] \\
= \frac12 \nabla_{\va} (\ln \det \Cova) +\frac12 \nabla_{\va}\tr \left[\Cova^{-1}N \right] \, .
\label{eq:meangradients}
\end{multline}
These contributions to the mean gradient are compensated by the second and third terms, respectively, in the loss function~\eqref{eq:lossfunc}.

The first term on the right of Eq.~\eqref{eq:meangradients} is nothing other than the mean-field of the standard MAP estimator. It corresponds to the mean-field induced by the delensing, and it is discussed in detail in \cite{Legrand:2025vzg}. The other term, $\nabla_{\va}\tr \left[\Cova^{-1} N \right]$, corresponds to subtracting from the delensing induced mean-field the part that comes directly from the delensing of the noise. Thus, our cross-only estimator is, in principle, insensitive to the mean-field created by the delensing of the noise.

\bibliography{biblio}

\end{document}